\documentclass[iop]{emulateapj}

\usepackage{graphicx}
\usepackage{longtable}
\usepackage{amsmath}

\newcommand{\RXS}{1RXS~J173006.4+033813}
\newcommand{\RXSshort}{1RXS~J1730+03}

\newcommand{\gpr}{\textit{g$\,^\prime$}}
\newcommand{\rpr}{\textit{r$\,^\prime$}}
\newcommand{\ipr}{\textit{i$\,^\prime$}}

\newcommand{\ha}{H$\alpha$}
\newcommand{\iraf}{\texttt{IRAF}}
\newcommand{\mdwarf}{M-dwarf}
\newcommand{\etal}{et.\,al }

\newcommand{\whathow}[1]{\noindent\textbf{What:} #1\\ \textbf{How/Why :} \noindent\begin{list}{$\bullet$}{\itemsep -1.3ex \topsep -1.3ex}}
\newcommand{\ewhathow}{\end{list}}
\newcommand{\figref}[1]{Figure \ref{fig:#1}}
\newcommand{\tabref}[1]{Table \ref{tab:#1}}
\newcommand{\eqnref}[1]{Equation (\ref{eq:#1})}
\shorttitle{The polar \RXSshort}
\shortauthors{Bhalerao \etal}
\submitted{ }
\begin{document}

\title{The polar Catalysmic Variable \RXS}

%\author{Varun B. Bhalerao, Fiona A. Harrison, Mansi Kasliwal, Shri Kulkarni, Vikram Rana}
%\author{Bhalerao, Varun B.\altaffilmark{a}; Harrison, Fiona A.\altaffilmark{b}; Kasliwal, Mansi M.\altaffilmark{a}; Kulkarni, Shrinivas R.\altaffilmark{a}; Rana, Vikram R.\altaffilmark{b}}
\author{Varun B. Bhalerao\altaffilmark{1}, Marten H. van Kerkwijk\altaffilmark{1,2}, Fiona A. Harrison\altaffilmark{3}, Mansi M. Kasliwal\altaffilmark{1}, S. R. Kulkarni\altaffilmark{1},Vikram R. Rana\altaffilmark{3}}
%\affil{California Institute of Technology, Pasadena, CA 91125, USA}

\altaffiltext{1}{Department of Astronomy, California Institute of Technology, Pasadena, CA 91125, USA}
\altaffiltext{2}{On sabbatical leave from Department of Astronomy and Astrophysics, 
University of Toronto, 50St. George Street, Toronto, ON M5S 3H4, 
Canada.}
\altaffiltext{3}{Space Radiation Laboratory, California Institute of Technology, Pasadena, CA 91125, USA}

\begin{abstract}
We report the discovery of \RXS, a polar cataclysmic variable with a period of 120.21\,min. The white dwarf primary has a magnetic field of $B = 42^{+6}_{-5}{\rm\,MG}$, and the secondary is a M3 dwarf. The system shows highly symmetric double peaked photometric modulation in the active state as well as in quiescence. These arise from a combination of cyclotron beaming and ellipsoidal modulation. The projected orbital velocity of the secondary is $K_2 = 390\pm4{\rm\,km\,s}^{-1}$. We place an upper limit of $830\pm65{\rm\,pc}$ on the distance.
\end{abstract}


\section{Introduction}\label{sec:intro}
Cataclysmic Variables (CVs) are close interacting binary systems in which a white dwarf (WD) accretes
material from a Roche lobe filling late-type secondary star \citep{cvbook,cvbook2}.
In most non-magnetic CVs ($B < 10^4{\rm\,G}$), the material lost from the secondary does not directly fall onto the WD because of its large specific orbital momentum: instead, it settles down in an accretion disc around the WD.

The accretion disc is the brightest component of the CV due to the large gravitational energy release in viscous accretion. The disc dominates the emission from the WD and donor over a wide wavelength range.

On the other hand, the accretion geometry in magnetic CVs is strongly influenced by the WD magnetic field. Magnetic CVs are broadly divided into two subclasses: Polars and Intermediate Polars (IPs). Polars usually show a synchronous or near synchronous rotation of WD with the orbital motion of the binary system and have high magnetic fields ($B > 10{\rm\,MG}$) \citep[for a review, see][]{polarreview}. In IPs the
WD rotation is far from synchronous and typically have magnetic field, $B < 10{\rm\,MG}$ \citep[for a review, see][]{ipreview}. The strong magnetic field in polars deflects the accretion material from a ballistic trajectory before an accretion disc can form, channeling it to the WD magnetic pole(s). The infalling material forms a shock near the WD surface, which produces radiation from X-rays to infrared wavelengths. Electrons in the ionized shocked region spiral around the magnetic field lines and emit strongly polarized cyclotron radiation at optical and infrared wavelengths. 
Polars exhibit X-ray on (high) and off (low) states more frequently than the other variety of CVs \citep{polaroff}.

\RXS\ (hereafter \RXSshort) is a Galactic source that is highly variable
in the optical and X-ray, exhibiting dramatic outbursts of more than 3
magnitudes in optical. It  was discovered by the ROSAT satellite during its all-sky
survey~\citep{rosat}. \citet{discovery}, in the course of their
investigation of poorly studied ROSAT sources, reported that USNO-B1.0
object \object[USNO-B1.0 0936-00303814]{0936-00303814} which is within the 10\arcsec\ (radius) localization
of the X-ray source showed great variability ($\Delta R$ of up to 3 mag)
in archival data (Palomar Sky Survey; SkyMorph/NEAT).  During certain
epochs the source appears to have been undetectable ($m_R > 20\,$mag).
\citet{discovery} undertook observations with Kazan State University's
30-cm robotic telescope and found variability on rapid timescales of 10
minutes.  

In this paper, we report the results of our photometric, spectroscopic and X-ray follow-up of \RXSshort.

%:----------------------
\section{Observations}\label{sec:obs}
\subsection{Optical photometry}\label{subsec:photometry}

We observed \RXSshort\ with the Palomar Robotic 60-inch telescope~\citep[P60;][]{cenko06} from UT 2009 April 17 to UT 2009 June 5, and with the Large Format Camera~\citep[LFC;][]{lfc} at the 5\,m Hale telescope at Palomar on UT 2009 August 26. Here we give details of the photometry.

We define a photometric epoch as observations from a single night when the source could be observed. We obtained 28 epochs with the P60, subject to scheduling and weather constraints. A typical epoch consists of consecutive 90--120\,s exposures spanning between 30--300 minutes (\tabref{photometry}). We obtained \gpr, \rpr, \ipr\ photometry on the first and third epochs.   After the third epoch, we continued monitoring the source only in \ipr\ band.

%: P60 photometry table - tab:p60photometry
\begin{deluxetable*}{cccccc}
\tablecaption{Photometry of \RXSshort\label{tab:photometry}}
\tablewidth{450pt}
\tablehead{
\colhead{Date (UT)} & \colhead{HJD} & 
\colhead{Filter name\tablenotemark{a}} & \colhead{Exposure time (sec)} &
\colhead{Magnitude} & \colhead{Error\tablenotemark{b}}
}
\startdata
% sample entries
20090417 & 54938.832278 &  \gpr  & 120 & 20.8 & 0.13\\
20090417 & 54938.835488 &  \ipr  & 90 & 19.39 & 0.05\\
20090417 & 54938.838696 &  \gpr  & 90 & 20.75 & 0.13\\
20090417 & 54938.839953 &  \ipr  & 90 & 19.17 & 0.05\\
20090417 & 54938.841212 &  \gpr  & 90 & 20.67 & 0.13\\
20090417 & 54938.842469 &  \ipr  & 90 & 19.1 & 0.05\\
20090417 & 54938.843727 &  \gpr  & 90 & 20.63 & 0.12\\
\nodata & \nodata   &  \nodata & \nodata & \nodata & \nodata \\
20090826 & 55069.691401 &  $lfci^\prime$  & 60 & 20.4 & 0.05\\
20090826 & 55069.693218 &  $lfci^\prime$  & 60 & 20.04 & 0.06\\
\enddata
\tablecomments{This table is available in a machine readable form online. A part of the table is reproduced here for demonstrating the form and content of the table.}
\tablenotetext{a}{Filters \gpr, \rpr\ and \ipr\ denote data acquired at P60 in the respective filters, $lfci^\prime$ denotes data acquired in the \ipr\ band with the Large Format Camera at the Palomar 200'' Hale telescope}
\tablenotetext{b}{Relative photometry error. Values do not include an absolute photometry uncertainty of 0.16\,mag in the \gpr\ band, 0.14\,mag in the \rpr\ band and 0.06\,mag in the \ipr\ band. Absolute photometry is derived from default P60 zero point calibrations.}
\end{deluxetable*}

%: Reference photometry table - tab:photrefs
\begin{deluxetable*}{cccccc}
%\rotate
\tablecaption{Photometry of reference stars for \RXSshort\label{tab:photrefs}}
\tablewidth{500pt}
\tablehead{
\colhead{Identifier\tablenotemark{a}} & \colhead{Right Ascension} & \colhead{Declination} &
\colhead{\gpr\ magnitude} & \colhead{\rpr\ magnitude} & \colhead{\ipr\ magnitude}
}
\startdata
% sample entries
A     & 262:30:21.99 & 03:38:37.5 & 15.861 $\pm$ 0.003 & 15.600 $\pm$ 0.003 & 15.341 $\pm$ 0.003 \\
B     & 262:30:14.57 & 03:37:11.0 & 17.372 $\pm$ 0.009 & 17.153 $\pm$ 0.006 & 16.872 $\pm$ 0.008 \\
C, 10 & 262:32:45.57 & 03:37:17.0 & 16.879 $\pm$ 0.006 & 16.691 $\pm$ 0.005 & 16.423 $\pm$ 0.006 \\
D, 9  & 262:32:13.84 & 03:38:03.0 & 18.161 $\pm$ 0.016 & 17.929 $\pm$ 0.010 & 17.660 $\pm$ 0.013 \\
E     & 262:31:00.31 & 03:38:31.3 & 18.518 $\pm$ 0.020 & 17.462 $pm$ 0.007 & 16.811 $pm$ 0.007 \\
F     & 262:30:55.06 & 03:37:26.4 & 18.530 $\pm$ 0.022 & 18.158 $pm$ 0.012 & 17.816 $pm$ 0.016 \\
G, 8  & 262:31:57.98 & 03:38:03.9 & 18.621 $\pm$ 0.022 & 18.259 $pm$ 0.012 & 17.905 $pm$ 0.016 \\
H     & 262:30:41.22 & 03:38:32.8 & 16.770 $\pm$ 0.006 & 16.081 $pm$ 0.004 & 15.688 $pm$ 0.004 \\
I     & 262:29:45.31 & 03:38:43.7 & 17.661 $\pm$ 0.011 & 17.302 $pm$ 0.007 & 16.969 $pm$ 0.008 \\
1     & 262:32:01.97 & 03:40:28.3 & \nodata & \nodata & 17.059 $\pm$ 0.035 \\
2     & 262:31:44.94 & 03:39:48.6 & \nodata & \nodata & 19.855 $\pm$ 0.043 \\
3     & 262:32:31.60 & 03:39:32.4 & \nodata & \nodata & 19.148 $\pm$ 0.039 \\
4     & 262:32:33.94 & 03:38:59.4 & \nodata & \nodata & 17.622 $\pm$ 0.033 \\
5     & 262:32:11.15 & 03:38:39.0 & \nodata & \nodata & 20.777 $\pm$ 0.049 \\
6     & 262:31:35.94 & 03:38:29.5 & \nodata & \nodata & 19.843 $\pm$ 0.026 \\
7     & 262:31:07.21 & 03:37:57.9 & \nodata & \nodata & 19.057 $\pm$ 0.026 \\
11    & 262:31:55.09 & 03:36:45.4 & \nodata & \nodata & 17.471 $\pm$ 0.030 \\
12    & 262:31:40.33 & 03:36:34.5 & \nodata & \nodata & 16.339 $\pm$ 0.044 \\
13    & 262:33:19.44 & 03:36:21.2 & \nodata & \nodata & 18.006 $\pm$ 0.030 \\
14    & 262:31:45.44 & 03:36:12.4 & \nodata & \nodata & 18.085 $\pm$ 0.048 \\
15    & 262:32:18.64 & 03:35:48.1 & \nodata & \nodata & 17.392 $\pm$ 0.055 \\
\enddata
\scriptsize{
\tablecomments{This table is available in a machine readable form online. A part of the table is reproduced here for demonstrating the form and content of the table.}
\tablenotetext{a}{The letters A--I denote stars used in photometry of P60 data, numbers 1--15 denote reference stars used in photometry of LFC images (\figref{finderwithrefs}, Section~\ref{subsec:photometry}).}
\tablenotetext{b}{Relative photometry error. Values do not include an absolute photometry uncertainty of 0.16\,mag in the \gpr\ band, 0.14\,mag in the \rpr\ band and 0.064\,mag in the \ipr\ band. Absolute photometry is derived using default P60 zero point calibrations.}
}

\end{deluxetable*}

\begin{figure*}[tbhp]
  \centering
  \includegraphics[height=185mm]{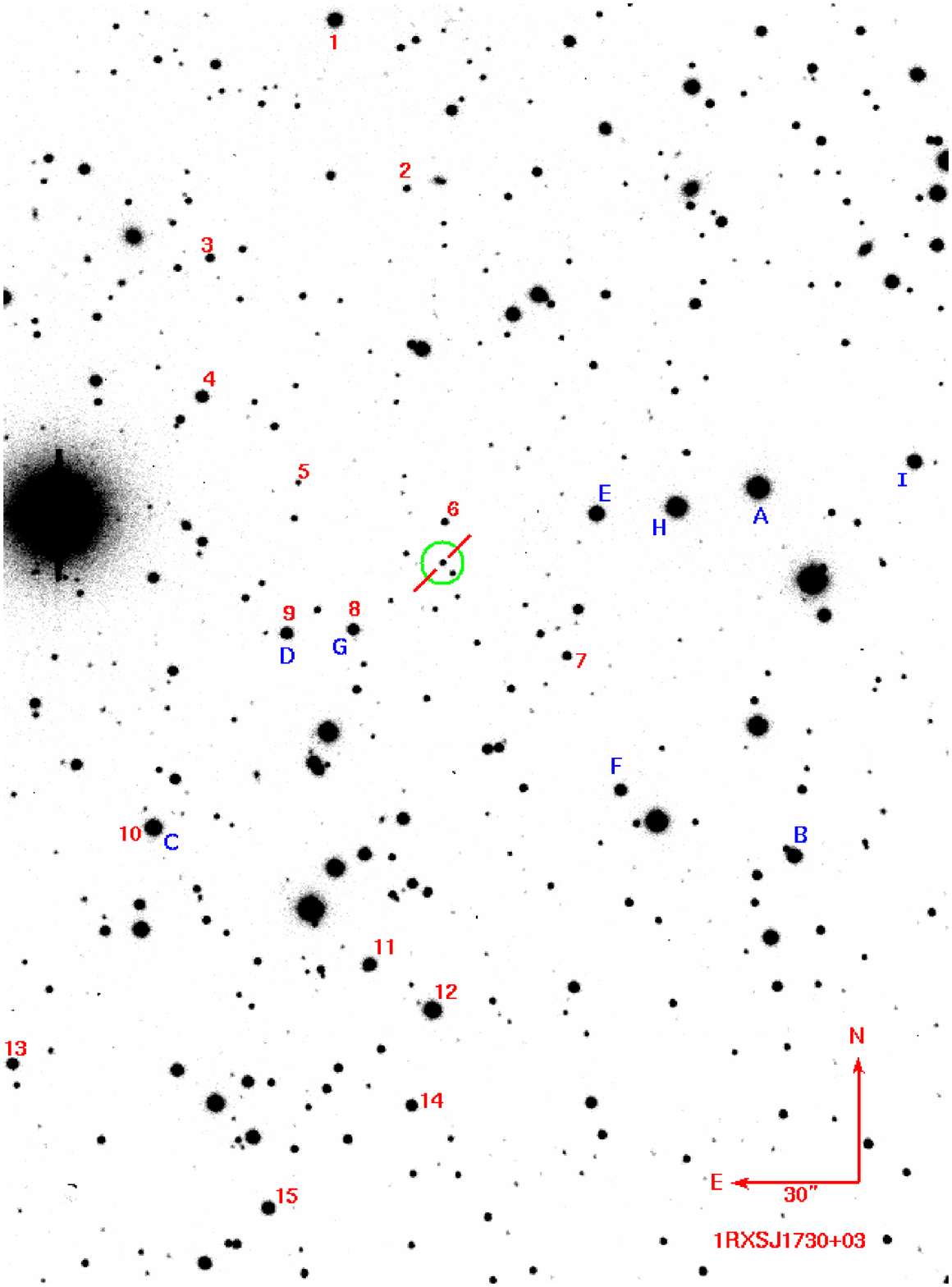} % requires the graphicx package
  \caption{Finder chart for \RXSshort\ ($\alpha = 17^{\rm h}$30$^{\rm m}$06$^{\rm s}$.19, $\delta= +03\degr$38\arcmin18\arcsec.8). This \ipr\ band image was acquired with the Large Format Camera (LFC) at the 5\,m Hale telescope at Palomar. Stars numbered 1--15 in red are 
  used for relative photometry in LFC data. Stars labeled \texttt{A}--\texttt{I} in blue are used for relative photometry in P60 data (Section\,\ref{subsec:photometry}). The green circle shows the 5\arcsec\ extraction region used for calculating UVOT fluxes. The circle includes a contaminator, 3\arcsec.4 to the South-West of the target.}
  \label{fig:finderwithrefs}
\end{figure*}

We reduced the raw images using the default P60 image analysis pipeline. LFC images were reduced in \iraf\footnote{http://iraf.noao.edu/}. We  performed photometry using the \texttt{IDL\footnote{http://www.ittvis.com/ProductServices/IDL.aspx} DAOPHOT} package~\citep{astrolib}. Fluxes of the target and reference stars (\figref{finderwithrefs}) were extracted using the \texttt{APER} routine. For aperture photometry, the extraction region was set to one seeing radius, as recommended by~\citet{aperturephot}. The sky background was extracted from an annular region 5--15 seeing radii wide. We used flux zero points and seeing values output by the P60 analysis pipeline. Magnitudes for the reference stars were calculated from a few images. The magnitude of \RXSshort\ was calculated relative to the mean magnitude of a 9 reference stars for LFC images, and 15 reference stars for P60 images (\tabref{photrefs}). The LFC images (\figref{finderwithrefs}) resolve out a faint nearby star ($m_{i^{\prime}} = 20.8$), 3\arcsec.4 from the target. The median seeing in P60 data is 2\arcsec.1 (Gaussian FWHM): so there is a slight contribution from the flux of this star to photometry of \RXSshort. We do not correct for this contamination. The statistical uncertainty in magnitudes is $\sim\,0.2\,$mag for P60 and $\sim\,0.05\,$mag for LFC, and the systematic uncertainty is  0.16\,mag in the \gpr\ band, 0.14\,mag in the \rpr\ band and 0.06\,mag in the \ipr\ band.

%: Finders

\begin{figure}[htbp]
\begin{center}
\includegraphics[scale=0.65, viewport=30 0 432 658]{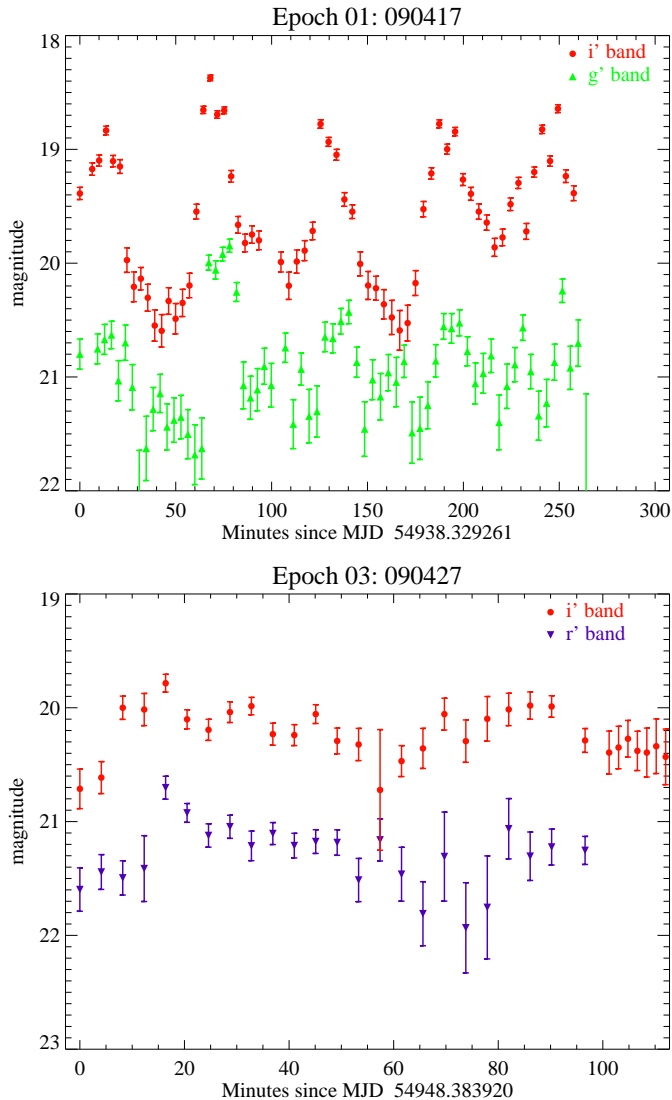}
\caption{P60 photometry of \RXSshort. Top panel: Epoch 1 photometry in \gpr\ and \ipr\ bands. Bottom panel: Epoch 3 photometry in \rpr\ and \ipr\ bands.}
\label{fig:p60photcompare}
\end{center}
\end{figure}

\begin{figure*}[htbp]
\begin{center}
\includegraphics[scale=0.8, viewport = 30 10 576 648]{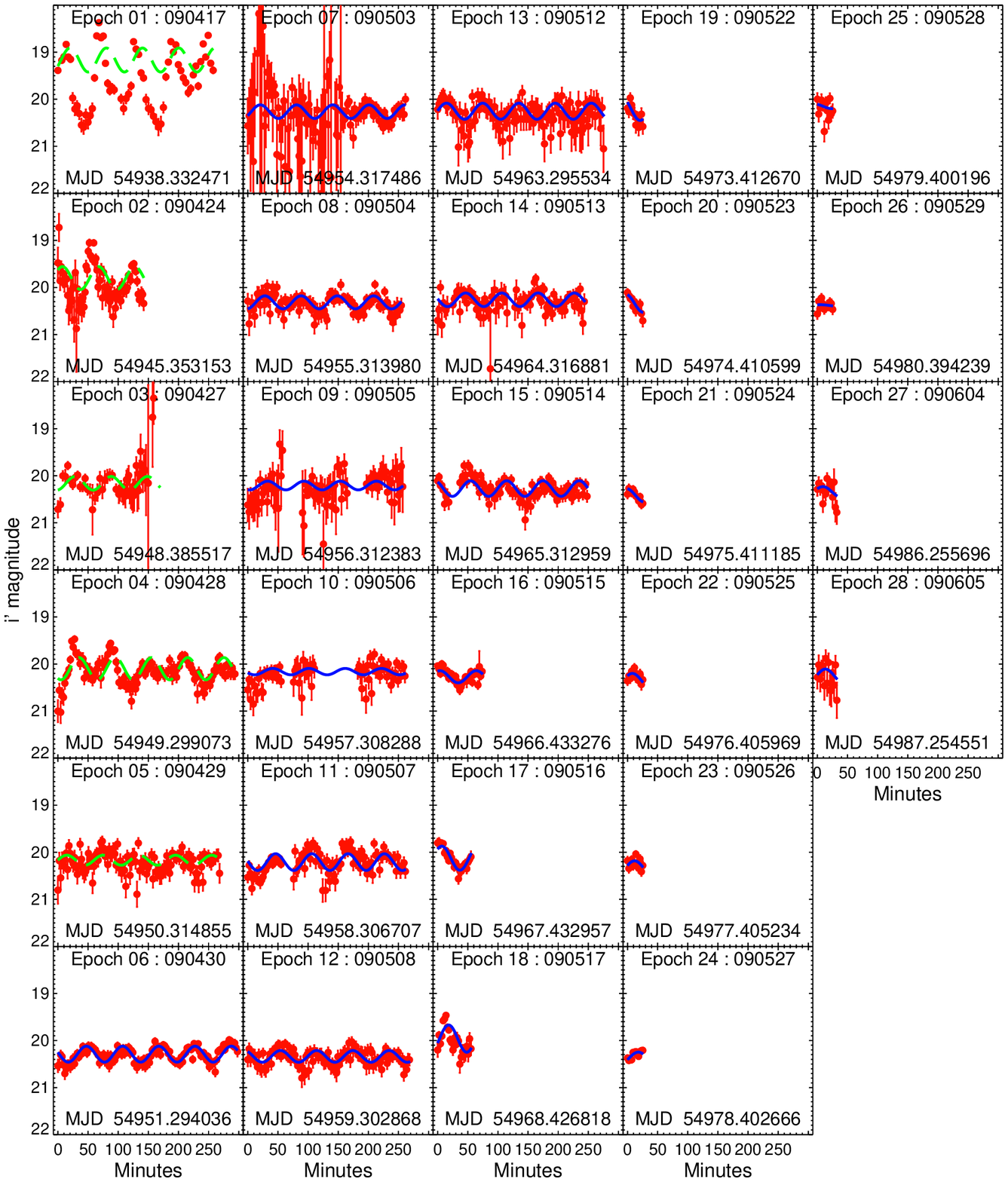}
\caption{\ipr\ band P60 photometry of \RXSshort, for Epochs 1 to 28. The source was in outburst in Epoch 1 and slowly faded into quiescence. The error bars denote relative photometry errors and do not include an absolute zero point error of 0.064\,mag (\S\,\ref{subsec:photometry}). The solid blue line denotes the best fit sinusoid to the photometric data. The dashed green line is added to show the relative phase of the source in outburst, these epochs were not used for fitting the lightcurve.}
\label{fig:p60phot1}
\end{center}
\end{figure*}

The resultant lightcurves are shown in Figures \ref{fig:p60photcompare}, \ref{fig:p60phot1} \& \ref{fig:lfcphot}.  \tabref{photometry} provides the photometry.

%: Swift data - (tab:swift)
\begin{deluxetable*}{ccccccccc}[thbp]
%\tabletypesize{\scriptsize}
%\rotate
\tablecaption{Observation Log for {\em Swift} ToO Observations of \RXSshort.\label{tab:swift}}
\tablewidth{500pt}
\tablehead{
\colhead{Obs ID} & \colhead{Start Date \& Time}  & \colhead{Stop Time} & 
\colhead{Exposure} & \colhead{Filter} & \colhead{Wavelength\tablenotemark{a}} & 
\colhead{Magnitude} & \colhead{Flux} & \colhead{Flux} \\
              &         &    & 
\colhead{(s)} &   & \colhead{(\AA{})} &
      &  & \colhead{($\mu$Jy)}
}

\startdata
00035571001 & 2006 Feb 9 16:56:43 & 18:42:00 & 1106 & UVM2 & 2231 & 17.9$\pm$0.1 & 31$\pm$2\tablenotemark{b} & 51$\pm$3 \\
& & & & XRT & & & 0.02\tablenotemark{c} & 0.06 (5\,keV) \\
00031408001 & 2009 May 3 18:28:56 & 19:02:12 & 1964 & UVM2 & 2231 & 20.44$\pm$0.21 & 3.04$\pm$0.59\tablenotemark{b} & 5.05$\pm$0.99 \\
00031408002 & 2009 May 4 18:34:04 & 22:12:54 & 4896 & UVW1 & 2634 & 20.14$\pm$0.10 & 3.47$\pm$0.32\tablenotemark{b} & 8.04$\pm$0.75 \\ 
00031408003 & 2009 May 6 03:02:21 & 15:36:33 & 5621 & UVW2 & 2030 & 21.48$\pm$0.17 & 1.38$\pm$0.22\tablenotemark{b} & 1.93$\pm$0.30 \\
& & & & XRT & & & $<$0.002\tablenotemark{c} & $<$0.006 (5\,keV) \\
\enddata

\tablenotetext{a}{Effective wavelength for each filter for a Vega-like spectrum~\citep{swiftvega}.} 
\tablenotetext{b}{Flux in the units of $10^{-17}$\,erg\,cm$^{-2}$\,s$^{-1}$\,A$^{-1}$. } 
\tablenotetext{c}{Counts\,s$^{-1}$ in 0.5-10\,keV.}

\end{deluxetable*}

%: P60 comparative photometry

%: P60 all i band photometry

\subsection{Spectroscopy}\label{subsec:spectroscopy}
We obtained optical and near-infrared spectra of \RXSshort\ at various stages after outburst (\figref{meanamplitude}). The first optical spectra were taken 13 days after the first photometric epoch. We used the Low Resolution Imaging Spectrograph on the 10\,m Keck-I telescope~\citep[LRIS;][]{lris}, with upgraded blue camera~\citep{lrisb1,lrisb2}, covering a wavelength range from 3,200\,\AA{}\,--\,9,200\,\AA{}. We acquired more optical data 34 days after outburst, with the Double Beam Spectrograph on the 5\,m Hale telescope at Palomar~\citep[DBSP;][]{dbsp}. We took 5 exposures spanning one complete photometric period, covering the 3,500\,\AA{}\,--\,10,000\,\AA{} wavelength range. We took late time spectra covering just over one photometric period for the quiescent source with the upgraded LRIS\footnote{http://www2.keck.hawaii.edu/inst/lris/lris-red-upgrade-notes.html}. At this epoch, we aligned the slit at a position angle of 45 degrees to cover both the target and the contaminator, 3\arcsec.4 to its South West (\figref{finderwithrefs}). We also obtained low resolution $J$-band spectra with the Near InfraRed Spectrograph on the 10\,m Keck-II telescope~\citep[NIRSPEC;][]{nirspec}. 12 spectra of 5 minutes each were acquired, covering the wavelength region from 11,500\,\AA{}\,--\,13,700\,\AA{}. For details of the
observing set up, see the notes to \tabref{vels}.

We analyzed the spectra using \iraf\ and \texttt{MIDAS}\footnote{Munich Image Data Analysis System; http://www.eso.org/sci/data-processing/software/esomidas/} and flux calibrated them using appropriate standards. Wavelength solutions were obtained using arc lamps and with offsets determined from sky emission lines. \figref{spectime} shows an optical spectrum from each epoch, while the IR spectrum is shown in \figref{irspec}.

The second LRIS epoch had variable sky conditions. Here, we extracted spectra of the aforementioned contaminator. This object is also a \mdwarf, hence both target and contaminator spectra will be similarly affected by the atmosphere. We estimate the \ipr\ magnitude of the contaminator in each spectrum, and compare it to the the value measured from the LFC images to estimate and correct for the extinction by clouds.

\subsection{X-ray and UV observations}\label{subsec:xray}

We observed the \RXSshort\ with the X-ray telescope (XRT) and the UV-Optical Telescope (UVOT) onboard the {\em Swift} X-ray satellite~\citep{swift} during UT 2009 May 3-6 for a total of about 12.5\,ks. The level-two event data was processed using {\em Swift} data analysis threads for the XRT (Photon counting mode; PC) using the \texttt{HEASARC FTOOLS\footnote{http://heasarc.gsfc.nasa.gov/ftools/;~\citet{ftools}}} software package. The source was not detected in the X-ray band. 

We follow the procedures outlined by~\citet{swiftvega} for analyzing the UVOT data. The measured fluxes are given in Table \ref{tab:swift}. The contaminator is within the recommended 5$^{\prime\prime}$ extraction radius. Hence, flux measurements are upper limits.

\citet{swiftatel} had observed \RXSshort\ on UT 2006 February 9 with the {\em Swift} satellite as a part of investigations of unidentified ROSAT sources. They detected the source with a count rate of 0.02${\rm\,counts\,s}^{-1}$. The best-fit power law has a photon index $\Gamma=1.8\pm 0.5$ and a 0.5--10\,keV flux of $1.2\times 10^{-12}\,{\rm erg\,cm^{-2}\,s^{-1}}$. After converting to the ROSAT bandpass assuming the XRT model parameters, this value is  approximately a factor of two lower than the archival ROSAT flux. They report a much higher UV flux in their observations, which suggests that the source was in an active state during their observations.

The column density inferred from the XRT data is low, $N_{\rm H}\sim 7\times 10^{20}\,$cm$^{-2}$. From ROSAT data, this column density corresponds to $A_{\rm V} = 0.39$, which gives $A_{\rm UVM2} = 0.84$~\citep{allens}. For comparison, \citet{extinction} give the Galactic dust extinction towards this direction ($l=26.^\circ7$, $b=19^\circ.7$) to be $E(B-V)=0.141$\,mag ($A_{\rm V} = 0.44$), corresponding to a column density of about $8\times 10^{20}\,{\rm cm}^{-2}$.

%: LFC photometry - fig:lfcphot
\begin{figure}[htbp]
\begin{center}
\includegraphics[scale=0.65, viewport=40 5 432 288]{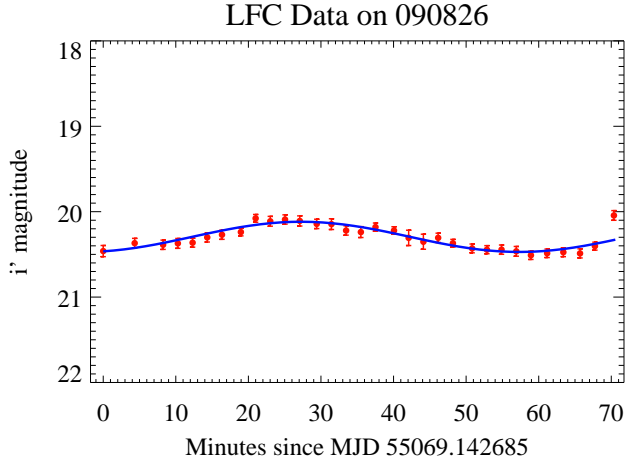}
\caption{\ipr\ band LFC photometry of \RXSshort\ in quiescence, 131 days after we first saw high variability. The variations have the same phase and period as observed during outburst. See Figure \ref{fig:p60phot1} for details.}
\label{fig:lfcphot}
\end{center}
\end{figure}

%:----------------------
\section{Nature of the components}\label{sec:components}
%\whathow{looks like a CV}
%\item red and blue components
%\item method for separating the components
%\ewhathow

\begin{figure*}[bthp]
\begin{center}
\includegraphics[height=70mm, viewport=50 20 648 432]{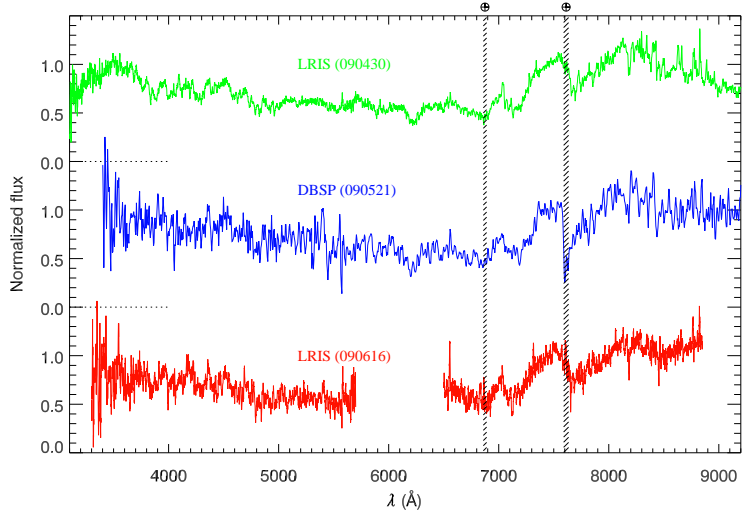}
\caption{Temporal evolution of \RXSshort\ spectra. Some spectra are not corrected for Telluric absorption, the affected regions are diagonally hatched. The first spectrum was obtained 13 days after we measured high variability, the second after 34 days, and the lowermost spectrum after 60 days. The spectra show broad Balmer features which evolve with time. The \mdwarf\ features (TiO, Na I) are clearly seen at all times.}
\label{fig:spectime}
\end{center}
\end{figure*}

The optical spectra (\figref{spectime}) show rising flux towards the red and blue ends of the spectrum: indicative of a hot (blue) and cool (red) component. The red part of the spectrum shows clear molecular features, characteristic of late type stars. The blue component is devoid of any prominent absorption/emission features. From the overall spectral shape we infer that \RXSshort\ is a CV.

\subsection{Red component}\label{subsec:mdwarf}

\begin{figure}[htbp]
\begin{center}
\includegraphics[scale=0.60,viewport=65 10 432 648]{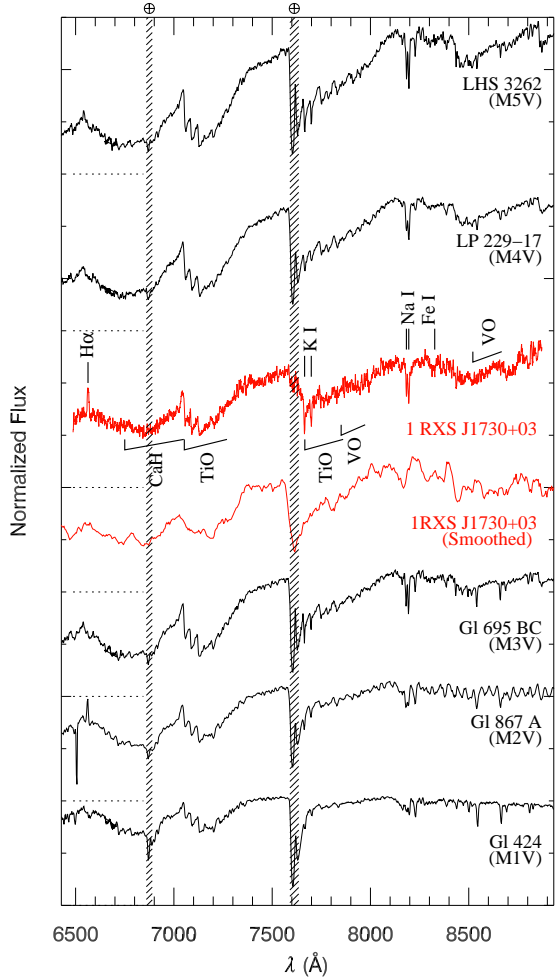}
\caption{Comparison of red part of \RXSshort\ spectrum with \mdwarf\ spectra. Some spectra are not corrected for Telluric absorption, the affected regions are diagonally hatched. Prominent bands (TiO, CaH) and lines (\ha, Na I) are marked. Comparing the shape of the TiO bands at 7053--7861~\AA{} and the shape of the continuum redwards of 8200~\AA{}, we infer that spectral type of the red component to be M3. The presence of a sodium doublet at 8183/8195\,\AA{} implies a luminosity class V.}
\label{fig:comparemdwarfs}
\end{center}
\end{figure}

The red component of \RXSshort\ is typical of a late type star. In \figref{comparemdwarfs}, we compare the red side spectrum of \RXSshort\ with several \mdwarf s. From the shape of the TiO bands at 7053\,\AA{}\,--\,7861\,\AA{}, we infer that the spectral type to be M3$\pm$1. This is consistent with the relatively featureless J band spectrum~\citep{bdss}. The spectral type indicates an effective temperature of 3400\,K \citep{allens}. The presence of a sodium doublet at 8183/8195\,\AA{} implies a luminosity class V.

We also fit the spectrum with model atmospheres calculated by \citet{spectralib}. For late type stars, these models are calculated in steps of $\Delta T =500{\rm\,K}$, $\Delta \log\,g=0.5$ and $\Delta {\rm [M/H]}=0.5$. We use model atmospheres with no rotational velocity $(V_{\rm rot}=0{\rm\,km\,s}^{-1})$ and convolve them with a kernel modeled on the seeing, slit size and pixel size. We ignore the regions contaminated by the telluric $A$ and $B$ bands (7615\,\AA{} and 6875\,\AA{}). The unknown contribution from the white dwarf was fit as a low order polynomial. We correct for extinction using $A_{\rm V}$ = 0.39 from X-ray data (Section\,\ref{subsec:xray}). To measure $\log\,g$, we use the spectrum in the 8,000\,\AA{}\,--\,8,700\,\AA{} region, which is expected to have fairly little contamination from the blue component. This region includes the \texttt{Ca II} lines at 8498, 8542\,\AA{} and the \texttt{Na I} doublet, which are sensitive to $\log\,g$. We then fit the spectra in the 6,700\,\AA{}\,--\,8,700\,\AA{} range to determine the temperature and metallicity. The best fit model has ${\rm T} = 3500{\rm\,K}$, $\log\,g = 5.0$ and solar metallicity, consistent with our determination of the spectral type.

\citet{donormass} state that unevolved donors in CVs follow the spectral type--mass relation of the zero age main sequence, as the effects of thermal disequilibrium on the secondary spectral type are negligible. For a M3 star, this yields a mass of $0.38\,M_{\odot}$. As the secondaries evolve, the spectral type is no longer a good indicator of the mass and gives only an upper limit on the mass. The lower limit can simply be assumed to be the Hydrogen-burning limit of $0.08\,M_{\odot}$.

%: Cyclotron humps (tab:cyclotron)
\begin{deluxetable}{cccc}
\tablecaption{Locations of Cyclotron Harmonics.\label{tab:cyclotron}}
\tablewidth{0pt}
\tablehead{
\colhead{Harmonic} & \colhead{Measured} & \colhead{Measured} &\colhead{Inferred} \\
\colhead{number} & \colhead{Wavelength}  & \colhead{Frequency} & \colhead{Wavelength\tablenotemark{a}} \\
 & \colhead{(\AA{})} & \colhead{(Hz)} & \colhead{(\AA{})}
}

\startdata
7 & 3540 & $8.4\times 10^{14}$ & 3664\\
6 & 4440 & $6.7\times 10^{14}$ & 4275\\
5 & 5180 & $5.7\times 10^{14}$ & 5130\\
4 & 6770 & $4.4\times 10^{14}$ & 6413\\
3 & 8225 & $3.6\times 10^{14}$ & 8551\\
2 & \nodata & \nodata & 12826\\
1 & \nodata & \nodata & 25653\\
%3540, 4440, 5180, 6767, 8221, 12900
\enddata
\scriptsize{
\tablenotetext{a}{Calculated from the best fit magnetic field strength, $B = 42{\rm\,MG}$}
}
\end{deluxetable}

\begin{figure}[thbp]
\begin{center}
\includegraphics[scale=0.6, viewport=30 0 432 648]{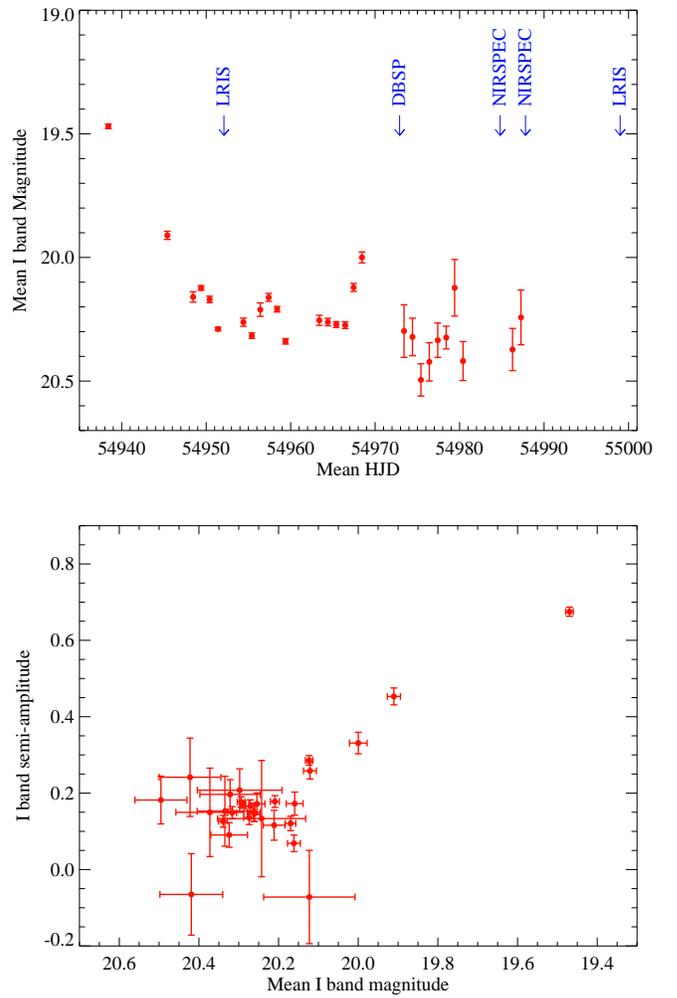}
\caption{Photometric evolution of \RXSshort\ in \ipr\ band. Data for each epoch was fit with a sinusoid with the same period for all epochs. In contrast to Figures \ref{fig:p60phot1} \& \ref{fig:lfcphot}, the phase was allowed to vary independently for each epoch. Top: average \ipr\ magnitude for each P60 observation epoch, as a function of time. Blue arrows mark spectroscopy epochs. Bottom: semi-amplitude of sinusoidal variations as a function of the corresponding mean \ipr\ magnitudes for each epoch.}
\label{fig:meanamplitude}
\end{center}
\end{figure}

%: Spectra

\subsection{Blue component}\label{subsec:whitedwarf}

The blue component of \RXSshort\ is consistent with a highly magnetic white dwarf. The blue spectrum is suggestive of a hot object, but does not show any prominent absorption/emission features. \ha\ is seen in emission, but other Balmer features are not detected. The spectrum (\figref{spectime}) shows cyclotron humps, suggesting the presence of a strong magnetic field. The polar nature of the object is supported by the absence of an accretion disc, and the transition from an active state to an off state in X-rays (Section\,\ref{sec:intro}, Section\,\ref{subsec:xray}).

%: IR spectrum
\begin{figure*}[bhtp]
\begin{center}
   \centering
   \includegraphics[scale=0.50, viewport=30 10 648 288]{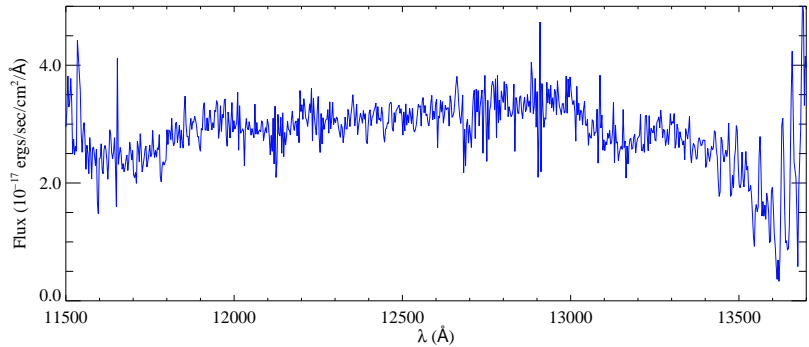}
   \caption{Keck NIRSPEC spectrum of \RXSshort. The spectrum is flat and nearly featureless, as expected for early M stars.}
   \label{fig:irspec}
\end{center}
\end{figure*}

%: Compare with other M dwarfs
\begin{figure*}[htbp]
\begin{center}
\includegraphics[scale=0.75, viewport = 20 0 648 484]{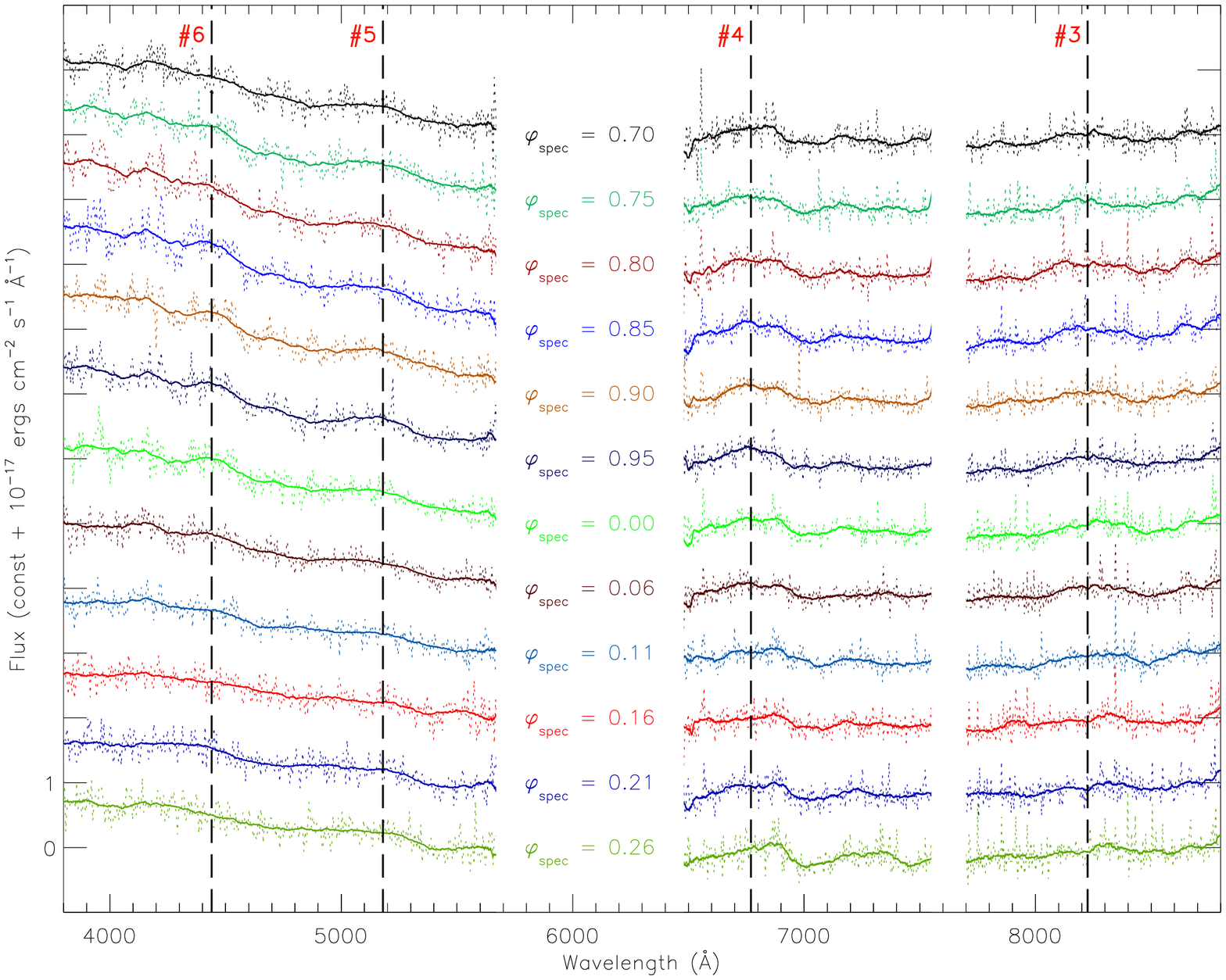}
\caption{Blue component spectra of \RXSshort, showing the 12 late-time LRIS exposures (Section\,\ref{subsec:spectroscopy}, \tabref{vels}). The spectroscopic phase at mid-exposure is indicated for each spectrum. Dotted lines show spectra are obtained by subtracting a scaled spectrum of GL\,694 from spectra of this binary. The overlaid solid lines are smoothed versions of the same spectra. Vertical dashed cyclotron humps at 4440\,\AA{}, 5180\,\AA{}, 6770\,\AA{} \& 8225\,\AA{}. The cyclotron harmonic numbers (\#3\,--\,\#6) are indicated in bold red. Two more cyclotron humps are seen in other spectra: a feature at 3540\,\AA{} (\figref{spectime}), and a $J$-band feature (\figref{irspec}).}
\label{fig:subtractmdwarf}
\end{center}
\end{figure*}

For analyzing the WD spectrum, we subtracted a scaled spectrum of the \mdwarf\ GL\,694 from the composite spectrum of \RXSshort. The resultant spectrum (\figref{subtractmdwarf}) clearly shows cyclotron harmonics. The hump seen in the J-band spectrum (\figref{irspec}) is also inferred to be a cyclotron harmonic. A detailed modeling of the magnetic field is beyond the scope of this work, but we use a simple model to estimate the magnetic field. We fit the cyclotron humps with Gaussians and measure the central wavelengths (\tabref{cyclotron}). We then fit these as a series of harmonics, and infer that the cyclotron frequency is $\nu_{\rm cyc} = 1.17\times10^{14}{\rm\,Hz}$. The magnetic field in the emission region is given by, $B  =  (\nu_{\rm cyc}/2.8\times10^{14}{\rm\,Hz}) \cdot 10^{8}{\rm\,G} = 42\times 10^6{\rm\,G}$.

%\begin{equation}\label{eq:magfield}
%B  =  \frac{\nu_{cyc}}{2.8\times10^{14}{\rm\,Hz}} \cdot 10^{8}{\rm\,G} = 41.7\times 10^6{\rm\,G}
%\end{equation}

As a conservative error estimate, we consider the worst case scenario where our identified locations for the cyclotron humps are off by half the spacing between consecutive cyclotron harmonics. Using this, we estimate the errors on the magnetic field: $B = 42^{+6}_{-5}{\rm\,MG}$.
%$B = 41.7 + 6.4 - 4.8$

%: Cyclotron humps

%:----------------------
\section{System parameters}\label{sec:thesystem}
\subsection{Orbit}\label{subsec:orbit}
%\whathow{Everything about the system orbit}
%\item Procedure for fitting RV
%\item orbital solution, K2, period,  mass function
%\item compare with other data points: keck, dbsp
%\ewhathow

We use the best fit \citet{spectralib} model atmosphere to measure radial velocities of the \mdwarf. We vary the radial velocity of the model, and minimize the $\chi^2$ over the 6,500\,\AA{}\,--\,8,700\,\AA{} spectral region, excluding the telluric O$_2$ bands. After a first iteration, the spectra are re-fit to account for motion of the \mdwarf\ during the integration time. For the 12 spectra taken at the second LRIS epoch, we also measure the radial velocity for the contaminator star on the slit, and find it to be constant. This serves as a useful test for our radial velocity measurement procedure. The barycentric corrected velocities are given in \tabref{vels}.

%convention: superior conjunction of primary is phase 0, i.e. when primary is furthest
% hence velocity will be mean and rising for m dwarf, i.e. sin(t-t0) as i have used

We fit a circular orbit ($v_2 = \gamma_2 + K_2 \sin([2\pi(t - t_0)]/P)$) to the measured velocities. We define the superior conjunction\footnote{When the WD is furthest from the observer along the line of sight.} of the WD as phase 0.

\begin{figure}[htbp]
\begin{center}
\includegraphics[scale=0.60, viewport = 20 0 432 576]{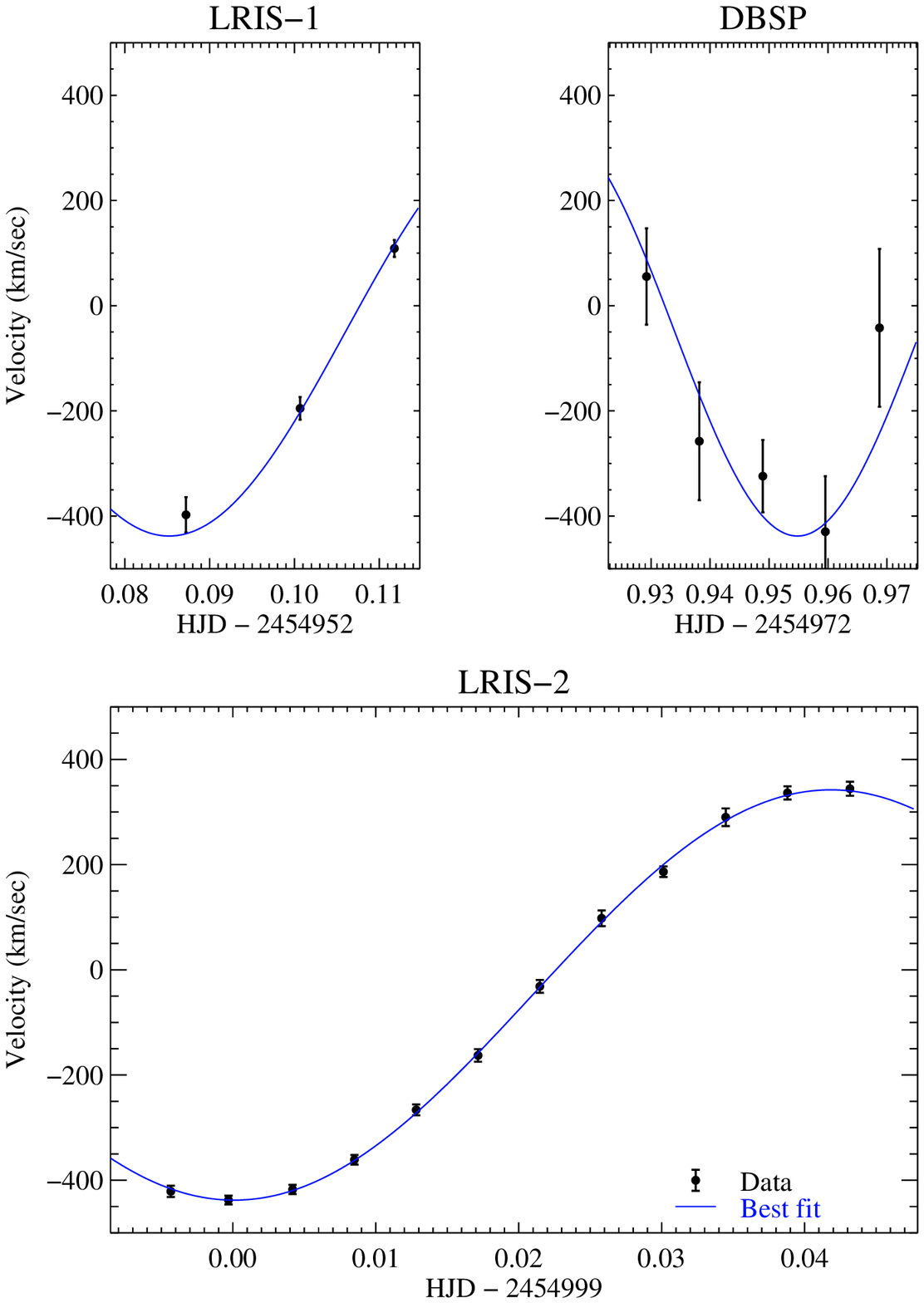}
\caption{Velocity measurements \RXSshort. The best-fit solution gives $P=120.2090\pm0.0013{\rm\,min}$, $\gamma_2=-48\pm5{\rm\,km\,s}^{-1}$ and $K_2 = 390\pm4{\rm\,km\,s}^{-1}$ (Section\,\ref{subsec:orbit})}
\label{fig:velfit}
\end{center}
\end{figure}

The 2009 June 16 spectroscopic data (\tabref{vels}) give an orbital period $P_{\rm est} = 123\pm3{\rm\,min}$. We then use the photometric variability (Section\,\ref{subsec:photvar}) to determine an accurate period in this range. Next, we refine the solution with velocity measurements from the other two spectroscopic epochs. The best-fit solution gives a period $P = 120.2090\pm0.0013{\rm\,min}$ and $K_2 = 390\pm4{\rm\,km\,s}^{-1}$ (\tabref{velfit}, \figref{velfit}).

%: velocity solution

%: h alpha moves
\begin{figure}[htbp]
\begin{center}
\includegraphics[scale=0.70, viewport = 50 0 360 504]{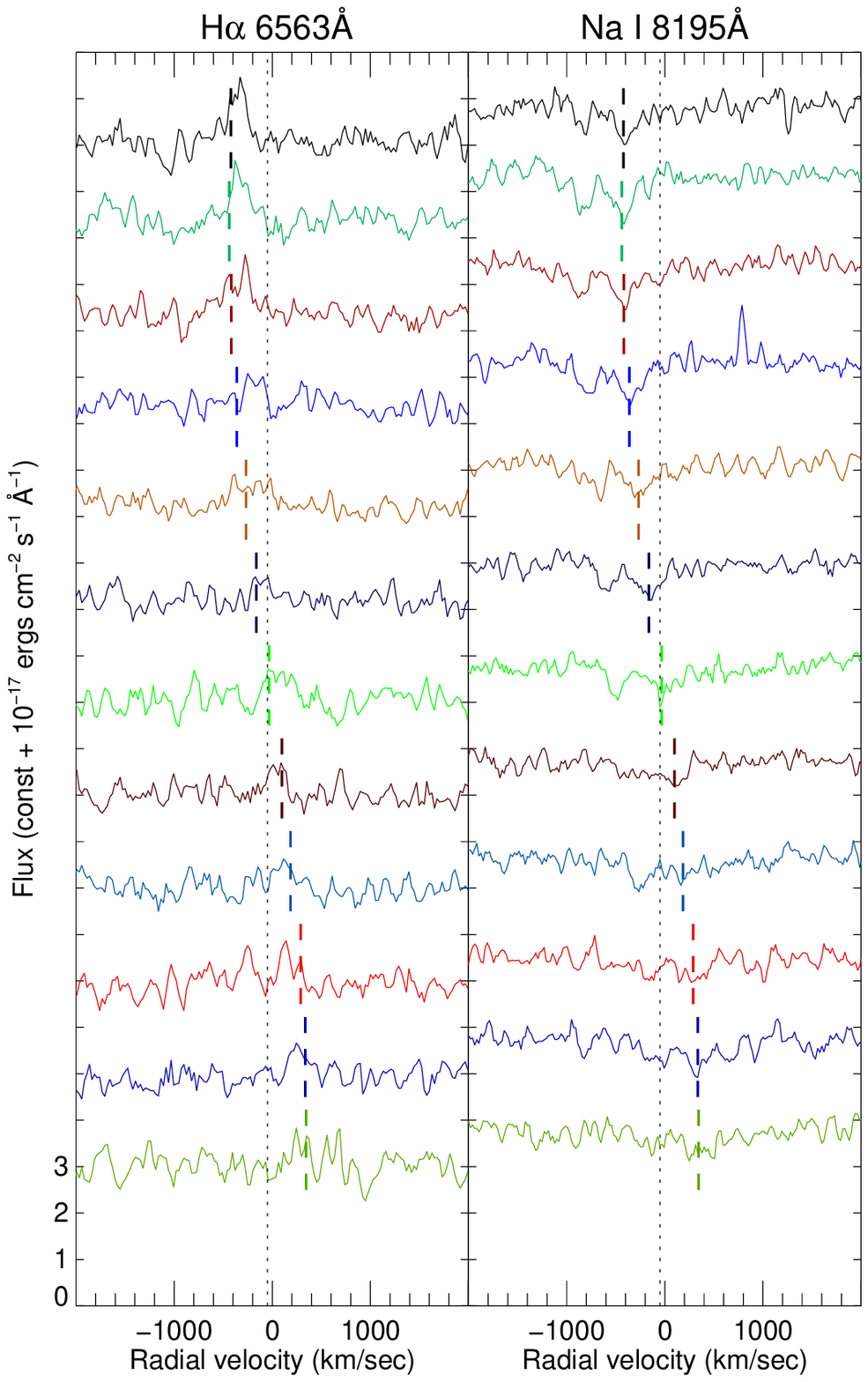}
\caption{Velocity modulation of \ha\ and the 8184/8195\,\AA{} \texttt{Na\,I} doublet. The spectra are offset by $2\times10^{-17}\,{\rm erg\,cm^{-2}\,s^{-1}}$ for clarity. The wavelengths are converted into velocities using the rest wavelengths 6563\,\AA{} and 8195\,\AA{} respectively. The vertical dotted line marks the radial velocity of the binary barycenter. The short dashed lines mark the radial velocities measured by fitting the complete spectrum. The \texttt{Na\,I} absorption follows the radial velocity of the \mdwarf, but \ha\ seems to have a smaller velocity amplitude.}
\label{fig:halpha}
\end{center}
\end{figure}

\figref{halpha} shows sections of the last epoch spectra around \ha\ and the \texttt{Na\,I} doublet at 8184/8195\,\AA{}. The \texttt{Na\,I} doublet clearly matches the velocity of the \mdwarf\ in the orbit, but the \ha\ emission seems to have a smaller velocity amplitude. A possible explanation for this is that the \ha\ emission comes from the \mdwarf\ surface that is closest to the WD, which may be heated by emission from the white dwarf or the accretion region.

%: Measured radial velocities (tab:vels)
\begin{deluxetable}{cccc}
\tablecaption{Radial velocity of the \mdwarf.\label{tab:vels}}
\tablewidth{0pt}
\tablehead{
 \colhead{Heliocentric JD} & \colhead{Exposure time} & \colhead{Barycentri} & \colhead{Instrument}\\
  &  & \colhead{Radial Velocity} & \\
 & \colhead{(s)} & \colhead{(${\rm\,km\,s^{-1}}$)}
}

\startdata
2454952.08721 & 1020 & $-398 \pm 34$ & LRIS\tablenotemark{a} \\
2454952.10067 & 1020 & $-195 \pm 22$ & LRIS\tablenotemark{a} \\
2454952.11179 & 600 & $109 \pm 16  $ & LRIS\tablenotemark{a} \\
2454972.92923 & 600 & $55 \pm 92   $ & DBSP\tablenotemark{b} \\
2454972.93818 & 900 & $-258 \pm 112$ & DBSP\tablenotemark{b} \\
2454972.94896 & 900 & $-324 \pm 69 $ & DBSP\tablenotemark{b} \\
2454972.95956 & 900 & $-430 \pm 105$ & DBSP\tablenotemark{b} \\
2454972.96875 & 600 & $-42 \pm 150 $ & DBSP\tablenotemark{b} \\
2454998.99566 & 210 & $-421 \pm 11 $ & LRIS\tablenotemark{c} \\
2454998.99969 & 300 & $-438 \pm 9  $ & LRIS\tablenotemark{c} \\
2454999.00418 & 300 & $-417 \pm 9  $ & LRIS\tablenotemark{c} \\
2454999.00850 & 300 & $-361 \pm 9  $ & LRIS\tablenotemark{c} \\
2454999.01283 & 300 & $-266 \pm 10 $ & LRIS\tablenotemark{c} \\
2454999.01715 & 300 & $-163 \pm 12 $ & LRIS\tablenotemark{c} \\
2454999.02147 & 300 & $-32 \pm 12  $ & LRIS\tablenotemark{c} \\
2454999.02580 & 300 & $98 \pm 15   $ & LRIS\tablenotemark{c} \\
2454999.03012 & 300 & $186 \pm 10  $ & LRIS\tablenotemark{c} \\
2454999.03448 & 300 & $290 \pm 17  $ & LRIS\tablenotemark{c} \\
2454999.03880 & 300 & $336 \pm 13  $ & LRIS\tablenotemark{c} \\
2454999.04317 & 300 & $344 \pm 13  $ & LRIS\tablenotemark{c} \\
\enddata

\tablenotetext{a}{~Low Resolution Imaging Spectrograph on the 10\,m Keck-I telescope~\citep{lris}, with upgraded blue camera~\citep{lrisb1,lrisb2}. Settings: Blue side: 3,200\,\AA{}\,--\,5,760\,\AA{}, grism with 400\,grooves/mm, blaze 3,400\,\AA{}, dispersion 1.09\,\AA{}\,pixel$^{-1}$, $R\,\sim\,$700. Red side: 5450\,\AA{}--9250\,\AA{}, grating with 400\,grooves/mm, blaze 8,500\,\AA{}, dispersion 1.16\,\AA{}\,pixel$^{-1}$, $R\,\sim\,$1,600. Dichroic: 5,600\,\AA{}, slit: 1\arcsec.0.}
\tablenotetext{b}{~Double Spectrograph on the 5\,m Hale telescope at Palomar~\citep{dbsp}. Settings: Blue side:  3,270\,\AA{}--5700\,\AA{}, grating with 600\,lines/mm, blaze 4,000\,\AA{}, dispersion 1.08\,\AA{}\,pixel$^{-1}$, $R\,\sim\,$1,300. Red side: 5,300\,\AA{}--10,200\,\AA{}, grating with 158\,grooves/mm, blaze 7,500\,\AA{}, dispersion 4.9\AA{}\,pixel$^{-1}$, $R\,\sim\,$600. Dichroic: D55 (5500\,\AA{}), slit: 1\arcsec.5.}
\tablenotetext{c}{~Low Resolution Imaging Spectrograph on the 10\,m Keck-I telescope, with upgraded blue and red cameras; http://www2.keck.hawaii.edu/inst/lris/lris-red-upgrade-notes.html. Settings: Blue side: 3,300\,\AA{}\,--\,5,700\,\AA{}, grism with 400\,grooves/mm, blaze 3,400\,\AA{}, dispersion 1.09\,\AA{}\,pixel$^{-1}$, $R\,\sim\,$700. Red side: 6,760\,\AA{}\,--\,8,800\,\AA{}, grating with 831\,grooves/mm, blaze 8200\,\AA{}, dispersion 0.58\,\AA{}\,pixel$^{-1}$, $R\,\sim\,$1,800. Dichroic: 5,600\,\AA{}, slit: 1\arcsec.0.}
\end{deluxetable}

%: Radial velocity solution (tab:velfit)
\begin{deluxetable}{rl}
\tablecaption{Orbital Velocity Parameters of the \mdwarf.\label{tab:velfit}}
\tablewidth{200pt}
\tablehead{
 \colhead{Parameter} & \colhead{Value}
}

\startdata
$\gamma_2$ (${\rm\,km\,s^{-1}}$)   &$-48   \pm 5$  \\
$K_2$ (${\rm\,km\,s^{-1}}$) &$390   \pm 4$  \\
$t_0$ (HJD)        &$54998.9375 \pm 0.0003$  \\
$P$ (min)          &$120.2090   \pm 0.0013$ \\
\enddata
\end{deluxetable}

\subsection{Photometric Variability}\label{subsec:photvar}

The lightcurve of \RXSshort\ shows clear periodicity (Figures \ref{fig:p60photcompare} -- \ref{fig:lfcphot}), with two peaks per spectroscopic period. Most of the photometric data was acquired in the \ipr\ band, which contains contribution from both the \mdwarf\ and a cyclotron harmonic from the emission region near the WD.

%: Joint Fourier transform
\begin{figure}[htbp]
\begin{center}
\includegraphics[scale=0.60, viewport=20 0 432 288]{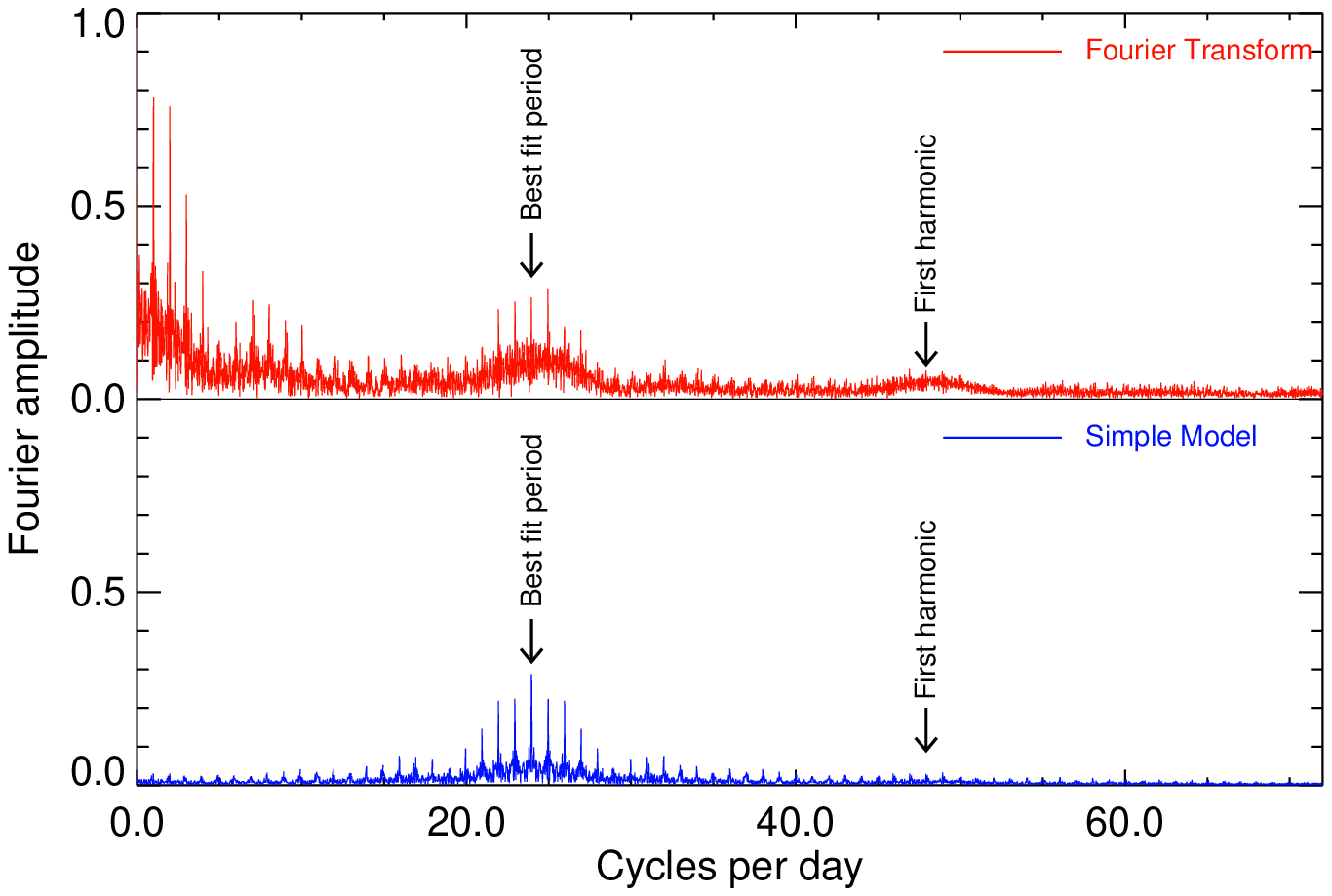}
\caption{Upper panel: Fourier transform of all \ipr\ data. Lower panel: expected Fourier transform for a pure sinusoidal variations with a period of 60.11\,min, obtained by scaling and convolving the Fourier transform of the window function with a delta function corresponding to the best-fit period.}
\label{fig:jointfourier}
\end{center}
\end{figure}

A Fourier transform of the data (\figref{jointfourier}) shows a strong peak at sixty minutes. We interpret this as a harmonic of the orbital period. To determine the exact period, we analyze the data as follows. As the object has a short orbital period, we convert all times to Heliocentric Julian Date for analysis. We fit a sinusoid ($m_0 + m_A \sin([2\pi(t - t_0)]/P)$) to each epoch, allowing $m_0$ and $m_A$ to vary independently for each epoch, but we use the same reference time $t_0$ and period $P$ for the entire fit. The mean magnitude is correlated with the amplitude (\figref{meanamplitude}). Note that the amplitude measured for the sinusoidal approximation for each epoch is always less than the actual peak-to-peak variations of the source during that epoch, as expected. The source is in the active state in the first few epochs, and we exclude epochs 1\,--\,5 from the fit, to avoid contamination from the accretion stream and/or the accretion shock.

The best-fit solution is overplotted in blue in Figures \ref{fig:p60phot1} \& \ref{fig:lfcphot}. Since epochs 1\,--\,5 were not included in the fit, a sinusoid is overplotted in dashed green to indicate the expected phase of the variations. The best-fit period is $60.1059\pm0.0005$ minutes. This formally differs from the spectroscopically determined orbital period by $2.1\sigma$. However, this error estimate includes only statistical errors. There is some, difficult to determine, systematic error component in addition, so we do not claim any significant inconsistency.

Periodic photometric variability for \RXSshort\ can be explained as a combination of two effects: cyclotron emission from the accretion region and ellipsoidal modulation. The active state is characterized by a higher mass transfer rate from the donor to the WD, and results in higher cyclotron emission. This emission is beamed nearly perpendicular to the magnetic field lines, creating an emission fan beam. For high inclination systems, the observer crosses this fan beam twice, causing two high amplitude peaks per orbit (\figref{p60photcompare}). In the active state when the emission is dominated by cyclotron radiation, the minimum leads the superior conjuntion by $\sim47$\degr.

In ellipsoidal modulation, and the photometric minima coincide with the superior conjunction. It is observed that in quiescence, the photometric minimum leads the superior conjunction of the white dwarf by $\sim14$\degr. This suggests that the $0.29\pm0.13$\,mag variation is caused by a combination of ellipsoidal modulation and cyclotron emission from the accretion region.

\subsection{Mass Ratio}\label{subsec:massratio}

The semi-amplitude of the M-dwarf radial velocity, K2, gives a lower limit on 
the mass of the white dwarf. For a circular orbit, one can derive from Kepler's laws that: 
\begin{equation}\label{eq:minwdmass}
M_{1, {\rm min}} = \frac{P K_2^3}{2\pi G} = 0.52\,M_{\odot}
\end{equation}
%\begin{equation}
%K_2 = v_2 \sin i = \left( \frac{2 \pi G [M_1 + M_2]}{P} \right)^{1/3} \sin i
%\end{equation}

Tighter constraints can be placed on the individual component masses $M_1$, $M_2$ by considering the geometry of the system. 
%The density of the secondary component depends on the period of the system: $\bar{\rho_2} =107 P^{-2}_{\rm hr}{\rm\,g\,cm^{-3}}$. 
\citet{rocheradius} expresses the volume radius $R_2$ of the secondary in terms of the mass ratio $q=M_2/M_1$: 
\begin{equation}\label{eq:r2}
\frac{R_2}{a} = \frac{0.49 q^{2/3}}{0.6q^{2/3} + \ln(1 + q^{1/3})}
\end{equation}
where the separation of the components is given by $a = [P^2 G(M_1 + M_2 )/4\pi^2 ]^{1/3}$ . Thus for a fixed period $P$, $R_2$ depends only on $M_2$. The radius of the white dwarf is much smaller than that of the \mdwarf. Hence, the \mdwarf\ will eclipse the accretion region if the inclination of the system is $i < \sin^{-1}(R_2/a)$. Figures \ref{fig:p60phot1}, \ref{fig:lfcphot} show that we do not detect any eclipses in the system.

\begin{figure*}[htbp]
\begin{center}
\includegraphics[scale=0.75]{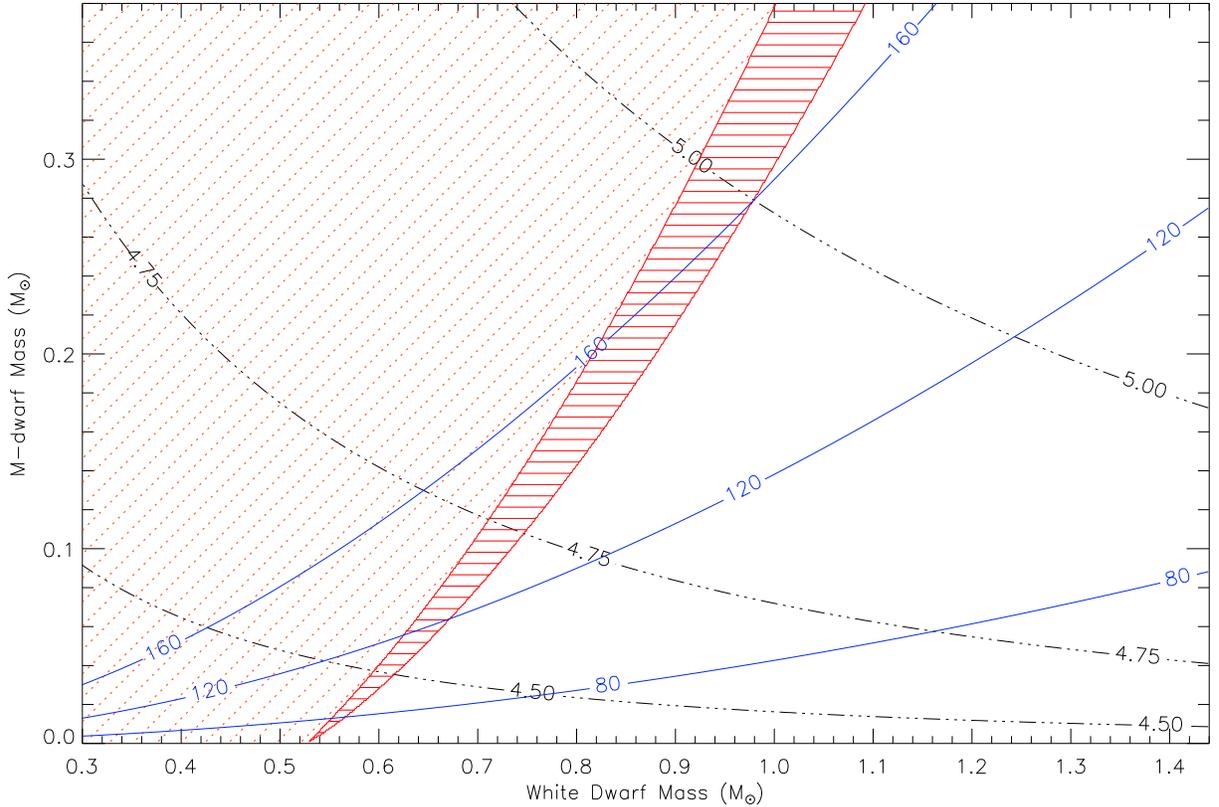}
\caption{The allowed range of masses (clear white region) for components of \RXSshort. The orange dotted region is excluded by the minimum inferred masses from radial velocity measurements. The red hashed region is excluded by non-detection of eclipses (Figures \ref{fig:p60phot1}, \ref{fig:lfcphot}). The black dash-dot lines show contours of constant $\log\,g$. The best fit spectra yeild $\log\,g = 5.0$. The solid blue lines are calculated contours for the rotation velocity (km\,s$^{-1}$) of the \mdwarf. A measurement of $v \sin i$ will help to constrain masses of the components.}
\label{fig:inclination}
\end{center}
\end{figure*}

%Given $M_1$ and $M_2$, there is a unique $i$ for the system such that $P$ and $K_2$ will have the measured values. 
\figref{inclination} shows the allowed region for \RXSshort\ in a WD mass -- \mdwarf\ mass phase space. The orange dotted region is excluded as the orbital velocity would be greater than the measured projected velocity. The red hatched region is excluded by non-detection of eclipses. The allowed mass of the primary ranges from the minimum mass ($M_1 > 0.52\,M_{\odot}$) to the Chandrashekhar limit. The mass of the secondary is bounded above by the ZAMS mass for a M3 dwarf ($M_2 = 0.38\,M_{\odot}$).
%: Mass-mass-inclination plot

For a given $M_2$ and a known orbital period, we can determine the radius of the secondary using \eqnref{r2}, and can calculate the surface gravity ($\log\,g$). For $0.38\,M_{\odot} \gtrsim M_2 \gtrsim 0.05\,M_{\odot}$, $\log\,g$ ranges from 5.1 to 4.8. This is consistent with $\log\,g = 5.0$ for the best fit \mdwarf\ spectrum (Section\,\ref{subsec:mdwarf}).

The donor stars in CVs are expected to co-rotate. For \RXSshort, the highest possible rotational velocity $v \sin i$ is $\sim\,160{\rm\,km\,s}^{-1}$ for a $0.38\,M_{\odot}$, $0.27\,R_{\odot}$ \mdwarf\ and a $1.1\,M_{\odot}$ WD (\figref{inclination}). $v \sin i$ will be lower if the WD is heavier or if the \mdwarf\ is lighter. To measure rotational broadening in the spectra, we use a higher resolution ($R=20,000$) template of the best-fit model from \citet{spectralib2}. We take a template with zero rotation velocity and broaden it to different rotational velocities using the prescription by \citet{rotbroaden}. Then we use our fitting procedure (Section\,\ref{subsec:mdwarf}) to find the best-fit value for $v \sin i$. For this measurement, we use only the 12 relatively high resolution spectra from the second LRIS epoch (UT 2009 June 16). The weighted $v \sin i$ from the twelve spectra is $97\pm22{\rm\,km\,s}^{-1}$, but the measurements show high scatter, with a standard deviation of $54{\rm\,km\,s}^{-1}$. We compared our broadened spectra with rotationally broadened spectra computed by \citet{spectralib2}, and found that our methods systematically underestimate $v \sin i$ by $\sim\,20{\rm\,km\,s}^{-1}$. We do not understand the reason for this discrepancy, hence do not feel confident enough to use this value in our analysis. A reliable measurement of $v \sin i$ will help better constrain the masses of the two components.

\subsection{Distance}

We estimate the distance to \RXSshort\ as follows. Our fitting procedure  (Section\,\ref{subsec:mdwarf}) corrects for extinction and separates the WD and \mdwarf\ components of the spectra. We correct for varying sky conditions by using a reference star on the slit. The ratio of measured flux to the flux of the best fit model atmosphere ($T = 3500{\rm\,K}, \log g = 5.0$) is,
\begin{equation}\label{eq:distanceratio}
\frac{f_{\rm measured}}{f_{\rm model}} = \left(\frac{R_2}{d}\right)^2 = (6.1\pm1.6)\times 10^{-23}
\end{equation}
where $d$ is the distance to the source, and $R_2$ is given by \eqnref{r2}.

For \RXSshort, the maximum mass of the \mdwarf\ is $0.38\,M_{\odot}$, and the corresponding radius is $R_2 = 0.27\,R_{\odot}$. \eqnref{distanceratio} then gives $d = 800\pm110{\rm\,pc}$. This calculation assumes the largest possible \mdwarf\ radius, hence is an upper limit to distance. If the \mdwarf\ is lighter, say $0.1\,M_{\odot}$, we get $R_2 = 0.17\,R_{\odot}$, yielding $d = 500\pm70{\rm\,pc}$.

\section{Conclusion}
\RXS\ is a polar cataclysmic variable, similar to known well-studied systems like BL Hyi, ST LMi and WW Hor in terms of the orbital period, magnetic field and variability between active and quiescent states. This source is notable for the highly symmetric nature and high amplitude of the double--peaked variation in the active state. This suggests a relatively high angle between the rotation and magnetic axes. Polarimetric observations of the source would help to better constrain the magnetic field geometry of the system.

Most polars are discovered due to their highly variable X-ray flux. However, we mounted a followup campaign for \RXSshort\ due to its unusual optical variability properties. This suggests that current and future optical synoptic surveys, such as PTF\footnote{http://www.astro.caltech.edu/ptf} \citep{ptf} and LSST can uncover a large sample of polars by cross-correlating opticaly variable objects with the ROSAT catalog.

%:----------------------
\section*{Acknowledgements}
We sincerely thank the anonymous referee for detailed comments on the paper. We thank N. Gehrels for approving the Target of Opportunity observation with {\em Swift}, and the {\em Swift} team for executing the observation. We also thank V. Anupama, L. Bildsten, T. Marsh, G. Nelemans, E. Ofek and P. Szkody and for useful discussions while writing the paper.

% dwarfarchives.org
This research has benefitted from the M, L, and T dwarf compendium housed at \verb+DwarfArchives.org+ and maintained by Chris Gelino, Davy Kirkpatrick, and Adam Burgasser. 

Some of the data presented herein were obtained at the W.M. Keck Observatory, which is operated as a scientific partnership among the California Institute of Technology, the University of California and the National Aeronautics and Space Administration. The Observatory was made possible by the generous financial support of the W.M. Keck Foundation.\\

{\it Facilities:} \facility{PO:1.5m}, \facility{Hale (LFC, DBSP)}, \facility{Keck:I (LRIS)}, \facility{Keck:II (NIRSPEC)}, \facility{Swift}

%:----------------------
%: Bibliography


\begin{thebibliography}

% ftools
\bibitem[Blackburn(1995)]{ftools} Blackburn, J.~K.\ 1995, 
Astronomical Data Analysis Software and Systems IV, 77, 367 

% P60 automation paper
\bibitem[Cenko et al.(2006)]{cenko06} Cenko, S.~B., et al.\ 
2006, \pasp, 118, 1396 

\bibitem[Cox(2000)]{allens} Cox, A.~N.\ 2000, Allen's 
Astrophysical Quantities, Springer, Berlin, 2000, 4 edn.

\bibitem[Cropper(1990)]{polarreview} Cropper, M.\ 1990, Space 
Science Reviews, 54, 195 

% discovery atel
\bibitem[Denisenko et al.(2009)]{discovery} Denisenko, D.~V., 
Kryachko, T.~V., 
\& Satovskiy, B.~L.\ 2009, The Astronomer's Telegram, 2014, 1 

% roche radius: eggleton
\bibitem[Eggleton(1983)]{rocheradius} Eggleton, P.~P.\ 1983, \apj, 
268, 368 


% swift satellite 
\bibitem[Gehrels et al.(2004)]{swift} Gehrels, N., et al.\ 
2004, \apj, 611, 1005 

% Stellar photospheres book: Gray. used for rotational broadening
\bibitem[Gray(2005)]{rotbroaden} Gray, D.~F.\ 2005, The 
Observation and Analysis of Stellar Photospheres, 3rd Edition, by 
D.F.~Gray.~ ISBN 
0521851866.~http://www.cambridge.org/us/catalogue/catalogue.asp? isbn=0521851866.~Cambridge, UK: Cambridge University Press, 2005.,  


%vikram
\bibitem[Hellier(2001)]{cvbook2} Hellier, C. 2001,
Springer-Praxis books in Astronomy \& Space Sciences, Chichester, UK  

%vikram
\bibitem[Hellier(2002)]{ipreview} Hellier, C. 2002, in The Physics of
Cataclysmic Variables and Related Objects, ed. B. T. G\"{a}nsicke, K.
Beuermann, \& K. Reinsch, ASP Conf. Ser., 261, 92

% Kolb: mass-temperature relation in CV donors
\bibitem[Kolb et al.(2001)]{donormass} Kolb, U., King, A.~R., 
\& Baraffe, I.\ 2001, \mnras, 321, 544 

% IDL
\bibitem[Landsman(1993)]{astrolib} Landsman, W.~B.\ 1993, 
Astronomical Data Analysis Software and Systems II, 52, 246 

% PTF paper
\bibitem[Law et al.(2009)]{ptf} Law, N.~M., et al.\ 2009, 
\pasp, 121, 1395 

% LRIS paper
\bibitem[McCarthy et al.(1998)]{lrisb1} McCarthy, J.~K., et 
al.\ 1998, \procspie, 3355, 81 

% BDSS - mclean et al brown dwarf spectroscopic survey. this is the lowres paper
\bibitem[McLean et al.(2003)]{bdss} McLean, I.~S., McGovern, 
M.~R., Burgasser, A.~J., Kirkpatrick, J.~D., Prato, L., 
\& Kim, S.~S.\ 2003, \apj, 596, 561 

% Nirspec paper
\bibitem[McLean et al.(1998)]{nirspec} McLean, I.~S., et al.\ 
1998, \procspie, 3354, 566 

% optimal radius for aperture photometry
\bibitem[Mighell(1999)]{aperturephot} Mighell, K.~J.\ 1999, 
Astronomical Data Analysis Software and Systems VIII, 172, 317 

% spectralib
\bibitem[Munari et al.(2005)]{spectralib} Munari, U., Sordo, R., Castelli, F., \& Zwitter, T.\ 2005, \aap, 442, 1127 

% DBSP paper
\bibitem[Oke 
\& Gunn(1982)]{dbsp} Oke, J.~B., \& Gunn, J.~E.\ 1982, \pasp, 94, 586 

% LRIS paper
\bibitem[Oke et al.(1995)]{lris} Oke, J.~B., et al.\ 1995, 
\pasp, 107, 375 


% Conversion from vega magnitudes for swift filters
\bibitem[Poole et al.(2008)]{swiftvega} Poole, T.~S., et al.\ 
2008, \mnras, 383, 627 

% polars switch off in x-rays:
\bibitem[Ramsay et al.(2004)]{polaroff} Ramsay, G., Cropper, M., 
Wu, K., Mason, K.~O., C{\'o}rdova, F.~A., 
\& Priedhorsky, W.\ 2004, \mnras, 350, 1373 

% Extinction maps
\bibitem[Schlegel et al.(1998)]{extinction} Schlegel, D.~J., 
Finkbeiner, D.~P., \& Davis, M.\ 1998, \apj, 500, 525 

% RXS swift atel
\bibitem[Shevchuk et al.(2009)]{swiftatel} Shevchuk, A.~S., Fox, 
D.~B., Turner, M., 
\& Rutledge, R.~E.\ 2009, The Astronomer's Telegram, 2015, 1 

% LFC paper
\bibitem[Simcoe et al.(2000)]{lfc} Simcoe, R.~A., Metzger, 
M.~R., Small, T.~A., 
\& Araya, G.\ 2000, Bulletin of the American Astronomical Society, 32, 758 

% LRIS paper
\bibitem[Steidel et al.(2004)]{lrisb2} Steidel, C.~C., 
Shapley, A.~E., Pettini, M., Adelberger, K.~L., Erb, D.~K., Reddy, N.~A., 
\& Hunt, M.~P.\ 2004, \apj, 604, 534 

% rosat survey
\bibitem[Voges et 
al.(1999)]{rosat} Voges, W., et al.\ 1999, \aap, 349, 389 

% Book about CVs
\bibitem[Warner(1995)]{cvbook} Warner, B.\ 1995, Cambridge 
Astrophysics Series, Cambridge, New York: Cambridge University Press, 
|c1995,  


\bibitem[Zwitter et 
al.(2004)]{spectralib2} Zwitter, T., Castelli, F., \& Munari, U.\ 2004, \aap, 417, 1055 


\end{thebibliography}
\end{document}